\documentclass[fleqn,10pt]{wlscirep}
\usepackage[utf8]{inputenc}
\usepackage[T1]{fontenc}
\usepackage{graphicx}
\usepackage{amsmath}
\usepackage{amssymb}
\usepackage{units}
\usepackage{algorithm}
\usepackage{algpseudocode}
\usepackage{subfigure}
\usepackage{pbox}
\usepackage{booktabs}       
\usepackage{array}
\newcolumntype{P}[1]{>{\centering\arraybackslash}p{#1}}
\usepackage{wrapfig}        

\title{Playing Ping Pong with Light:\\ Directional Emission of White Light}

\author[1,2,*]{Heribert Wankerl}
\author[2]{Christopher Wiesmann}
\author[2]{Laura Kreiner}
\author[2]{Rainer Butendeich}
\author[2,3,4]{Alexander Luce}
\author[2]{Sandra Sobczyk}
\author[2]{Maike Lorena Stern}
\author[1]{Elmar Wolfgang Lang}
\affil[1]{University of Regensburg, 93053 Regensburg, Germany}
\affil[2]{OSRAM Opto Semiconductors GmbH, 93055 Regensburg, Germany}
\affil[3]{Max Planck Institute for the Science of Light,  91058 Erlangen, Germany}
\affil[4]{Friedrich Alexander University Erlangen Nuremberg, 91054 Erlangen, Germany}
\affil[*]{heribert.wankerl@ams-osram.com}


\begin{abstract}
Over the last decades, light-emitting diodes (LED) have replaced common light bulbs in almost every application, from flashlights in smartphones to automotive headlights. Illuminating nightly streets requires LEDs to emit a light spectrum that is perceived as pure white by the human eye. The power associated with such a white light spectrum is not only distributed over the contributing wavelengths but also over the angles of vision. For many applications, the usable light rays are required to exit the LED in forward direction, namely under small angles to the perpendicular. In this work, we demonstrate that a specifically designed multi-layer thin film on top of a white LED increases the power of pure white light emitted in forward direction. Therefore, the deduced multi-objective optimization problem is reformulated via a real-valued physics-guided objective function that represents the hierarchical structure of our engineering problem. Variants of Bayesian optimization are employed to maximize this non-deterministic objective function based on ray tracing simulations. Eventually, the investigation of optical properties of suitable multi-layer thin films allowed to identify the mechanism behind the increased directionality of white light: angle and wavelength selective filtering causes the multi-layer thin film to play ping pong with rays of light.
\end{abstract}
\begin{document}

\flushbottom
\maketitle
%
%
\thispagestyle{empty}

\section{Introduction}
\label{introduction}
\begin{figure}[t]
\vspace{-1.0cm}
\subfigure[Conceptual LED setup]{\includegraphics[width=0.49\textwidth]{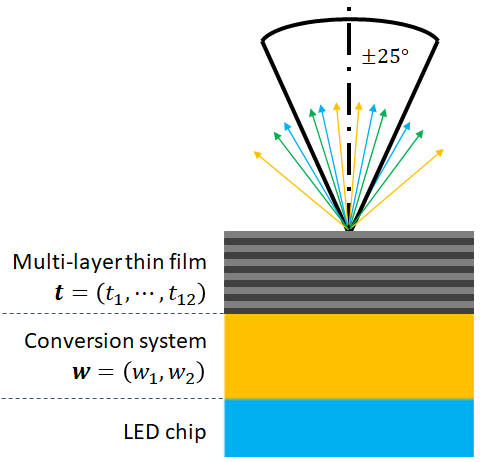}}
\subfigure[Color space according to the \textbf{C}ommission \textbf{i}nternationale de l’\textbf{é}clairage (CIE-color space adopted from Ding et al. \cite{Ding2013})]{\includegraphics[width=0.49\textwidth]{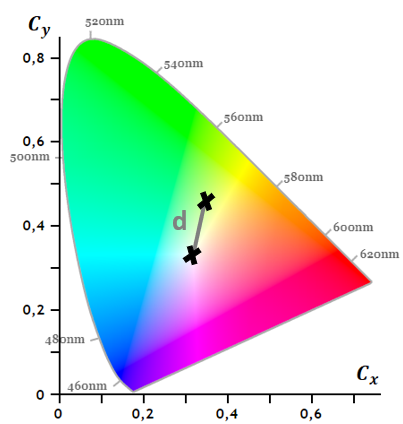}}
\caption{An LED (a) consists of a chip, a conversion system and a multi-layer thin film that features ten layers. The latter focuses the emitted light spectrum into a $\pm 25^\circ$-angle cone. The spectrum is composited of light rays (arrows) of different wavelengths. Such a spectrum can be associated with a point $\mathbf{c}=\left( C_x, C_y \right)$ in the color space (b). In this work, the Euclidean distance between two color points is denoted as the color point deviation $d$.}
\label{fig:introduction}
\end{figure}
For many recent applications in our everyday lives, light-emitting diodes (LED) are required to emit a light spectrum that features specific optical properties. Such a light spectrum explains how the (relative) power provided by an LED is distributed over its contributing wavelengths and occurring emergent angles. In this work, we elaborate on how to shift the light distribution of white LEDs towards forward directions using a multi-layer thin film (MLTF). As illustrated in figure \ref{fig:introduction} $(a)$, the term white LED (package) refers to a horizontal stack of a semi-conductor chip, the conversion system and an optional multi-layer thin film. Further on, the common LED structure without an MLTF is referred to as the reference design. By multiplication with a color matching function\cite{Guild1931,Wright1929}, each light spectrum can be characterized by the so-called color point, a two-dimensional vector in the color space of figure \ref{introduction} $(b)$, which describes the color of a spectrum\cite{Ding2013}. In the common case of a white LED without a multi-layer thin film (MLTF), the chip emits blue light that passes through the ensuing conversion system\cite{Cho2017}. Here, two conversion materials, compounded with weight percentages $\mathbf{w}=(w_1, w_2)$, convert a portion of the blue light to green and red light resulting in white light. The amount of conversion materials determines the degree of conversion and is adapted such that the resulting light spectrum corresponds to the application-specific white target color point $\textbf{C}$. Another important characteristic of an LED is the power of its radiated spectrum. In radiometry, this power is commonly referred to as \textit{radiant flux}. Hence, both the \textit{color point} $\mathbf{c}^\alpha(\mathbf{w})$ and the \textit{forward power} $P^\alpha(\mathbf{w})$ featured by an LED in a $\pm \alpha$-angle cone are computed based on the anisotropic radiated light spectrum. Unfortunately, such emitted spectra of LED packages are unknown a priori for varying conversion materials and/or MLTFs. Thus, the deduction of spectrum-related measures of a particular LED like its power or color point would require expensive physical experiments; namely, to physically fabricate and evaluate the spectra of such LED packages. An elegant and cheaper alternative is to statistically trace a bunch of rays sampled from the light distribution of the light injecting LED chip, until they exit the LED package and form the emitted spectrum, or vanish due to non-radiative thermal losses. Thereby, the behavior of each ray is governed by geometrical optics\cite{Lahiri2016}. In this framework, only the spectrum of the LED chip needs to be known and thus is measured once. The spectrum of the LED chip in turn is assumed to remain stable over changes of both, the conversion system and the MLTF. Such still time-consuming and noisy ray tracing simulations are based on calibrated optical models of LEDs \cite{Kenji2018,Liu2010,Sun2008,Lee2001,Sun2006} and allow an estimation of their spectra. Notably, ray tracing simulations are non-deterministic and can be considered Monte-Carlo simulations \cite{Lee2001}: The random variables of interest (power and color point) cannot be computed analytically and thus need to be estimated via numerous realizations and processing (tracing) of observable random variables (rays that exit the chip). Such tray tracing simulations replace expensive experiments and allow to conduct extensive optimization of LEDs.\par
Since the aforementioned so-called \textit{color point optimization problem} based on $\mathbf{w}$ is empirically unambiguous and convex, it is not possible to further increase the power $P^\alpha(\mathbf{w})$ if
\begin{align*}
\mathbf{w} = \text{argmin}_{\boldsymbol{\omega}} \lbrace d \left( \mathbf{C} , \textbf{c}^\alpha\left( \boldsymbol{\omega} \right) \right) \rbrace
\end{align*}
holds based on the Euclidean distance $d \equiv \Vert \cdot \Vert_2$. The proposed idea of this manuscript is to modify the anisotropic light spectrum so as to focus more power $P^\alpha(\mathbf{w}, \mathbf{t})$ into the forward angle cone compared to the reference design by using an MLTF --- parameterized by $\mathbf{t}$ --- on top of the conversion system; thus, increasing the directionality of white light. In this work, directional emission or directionality of white light refers to the power that is emitted in a particular (solid) angle of interest induced by emitted rays of light accumulated in these directions. The corresponding inverse design problem features $T + 2$ parameters that describe the modified LED: In addition to the weight percentages of the conversion materials, the layer thicknesses $\mathbf{t} = \left( t_1, ..., t_{T} \right)$ of $T$ alternating layers of titanium dioxide ($\text{TiO}_2$) and silicon dioxide ($\text{SiO}_2$) can be adapted. Notably, increasing the usable power may compete with achieving the application-specific white target color point. For instance, due to the \textit{Stokes shift} \cite{Stokes1852}, designing an MLTF that raises the ratio of blue light in forward direction, obviously increases the usable power but will no longer retain the desired color point. Namely, it will render the LED's light to appear bluish and no longer convenient for a white light application. In other words, the MLTF is required to not only increase the number of rays that emerge in forward direction but also achieve a particular ratio between blue, green and red rays. The aforementioned challenges result in a multi-objective optimization problem: improving the directionality of white light while preserving the color point associated to the spectrum that exits the LED. More precisely, we aim to increase $P^\alpha(\mathbf{w}, \mathbf{t})$ while keeping the Euclidean distance $ d^\alpha(\mathbf{w}, \mathbf{t}) \equiv d(\mathbf{c}^\alpha(\mathbf{w}, \mathbf{t}), \mathbf{C})$ low. To summarize, the contribution of this work is three-fold:
\begin{itemize}
\item Bayesian optimization is used to adapt the layer thicknesses of an MLTF based on ray tracing simulations of white LEDs
\item MLTFs are investigated with regard to their general ability to increase directionality of white light emission of LEDs
\item The effect that explains the directionality increase of white LEDs in physical terms is identified
\end{itemize}

\section{State of the art}
Although the directional emission of white light is crucial for many applications like head lamps of cars, to the best of our knowledge, none of the aforementioned points were investigated in previous work. Traditionally, the directionality of incoherent light sources like LEDs is increased via optical devices such as reflectors or lenses \cite{Lorenzo1982,Ma2015}, which shape the beam profile of a light source \cite{Fournier2011}. Notably, such optics are often bulky and therefore cause optical loss or limit the design. On the nanoscale, meta-\cite{Su2020,Lalanne2017, Schreiber2003}
and microlenses\cite{Lee2013,Chen2018} have been introduced to manipulate light by microscopic structures. In addition, scientific work has been conducted towards the development of incoherent and directional light sources based on periodic or random gratings like dielectric\cite{Yang2017,Yan2014} or plasmonic\cite{Kamakura2018} structures. Moreover, extensive studies regarding metasurfaces \cite{Agata2021,Shunsuke2021} have been conducted to increase the directional outcoupling of emitters. However, all of these systems are challenging to fabricate in mass-production or apply to coherent light sources only. Notably, they do not affect the light source itself but rather shape the light distribution over angle that exits a light source like an LED package. In contrast, the proposed epitaxial deposition of an additional MLTF on top of the conversion system of an LED is straightforward from an engineering viewpoint, allows compact integration in existing LED packages, implies only low additional expenditure and is directly applicable for mass-production of optical semiconductors. The closest investigation to ours may be the study conducted by Yi Zheng and Matthew Stough \cite{Zheng2008}: They proposed to use an MLTF as a wavelength selective filter between the LED chip and the conversion system to increase the global efficacy of white LEDs. This filter allows the blue light coming from the chip to pass while simultaneously reflecting the green and red light emitted by the conversion system. Thereby, reabsorbtion effects of non-blue photons in the chip are suppressed. In other words, the MLTF enforces the green and red light to exit the package rather than irradiating the chip and thus increases the overall extraction efficiency. Such an MLTF is included into the optical LED model used in this work and further considered as integral component of the LED chip. In contrast, the contribution of our work is to improve the directional emission into an angle cone rather than increasing the overall outcoupling efficiency. Therefore, we demonstrate that an MLTF, acting as an angle and wavelength selective filter between the conversion system and the ambient air, can increase the directional emission of white light. Thereby, the MLTF helps to shape the angular and spectral light distribution directly through ray ping pong during its generation process rather than to collimate the outcoupled light.\par
Designing MLTFs that feature a particular target reflectivity or transitivity over angle of incidence and wavelength is a common engineering challenge and therefore many optimization methods have been developed: Some of them are based on gradients \cite{Sullivan1996} or biological inspirations \cite{rabady2014,Guo2014}.
Recently, even some approaches including neural networks \cite{Roberts2018,LiuII2018,Hedge2019} or reinforcement learning \cite{Wankerl2021,Jiang2020} have been implemented to efficiently scan the search space for suitable MLTF designs. These techniques rely on fast-to-evaluate computations regarding the transfer matrix method\cite{Luce2021,Byrnes2020} (TMM) that allow to compute the optical characteristics\footnote{e.g. reflectivity or transmittivity over wavelength and angle of incidence} of an MLTF. To measure\footnote{e.g. with a notion of reconstruction error} how close a particular MLTF's optical characteristic is to an optimal one takes not even a second in total. However, in contrast with applications like anti-reflection coatings\cite{Guo2014}, a specific optimal optical characteristic that causes an MLTF to increase the directionality of white light is not known a priori. To circumvent this lack of information, we conduct noisy ray tracing simulations in order to optimize the power and color point in forward direction regarding the layer thicknesses of an MLTF. As a side benefit, the optical characteristic of MLTFs is implicitly optimized towards the a priori unknown optimal one and can be investigated further a posteriori. Notably, these noisy simulations take about $4 \text{ [min]}$ for a given MLTF, which is $420$ times longer compared to computing the optical characteristics of the MLTF itself based on TMM. Thus, ray tracing simulations are relatively expensive-to-evaluate and render most of the aforementioned data-hungry TMM-based optimization methods of MLTFs impractical. Therefore, we propose to adapt the individual MLTF layer thicknesses in order to maximize $P^\alpha(\mathbf{w}, \mathbf{t})$ while minimizing $d^\alpha(\mathbf{w}, \mathbf{t})$ via a variant of Bayesian optimization\cite{Bradford2018}. Although Bayesian optimization has shown satisfying results on many mathematical test functions as well as expensive-to-evaluate real-world problems regarding engineering, physical and chemical sciences \cite{Schweidtmann2018,Amar2019,Clayton2020}, to the best of our knowledge no application towards ray tracing simulations is reported in scientific literature. After optimizing an MLTF, its optical characteristics, like transmittivity, allow to physically deduce target behaviors that an MLTF needs to fulfill in order to further increase the directionality of white light. Hereby, an effect called \textit{ray ping pong} is identified to be responsible for the improved directional emission of white light. Related effects like photon recycling \cite{Cheng2021} were studied in previous work for solar GaAs cells \cite{Kosten2014,Raja2021,Walker2015}, perovskite LEDs \cite{Cho2020} or thin films themselves \cite{Fu2021}. In our work, depending on their wavelength and incident angle, the rays are considered to be partially trapped in the LED package by an MLTF, enforcing interactions between rays and LED package materials like scattering, (re-)absorbtion, (re-)emission or non-radiative effects. The MLTF can statistically steer the degree of such interactions for particular parts of the original spectrum emitted by the LED package. Thus, the MLTF has a direct impact on the emitted spectrum and needs to balance the radiated spectrum not only to appear as white in forward direction, but also to suppress radiation in non-forward direction. In other words, the MLTF is found to play angle and wavelength selective ping pong with the rays of light to achieve an equilibirium of emitted rays, which is of advantage regarding directionality while still holding the color point.

\begin{figure}[!t]
\vspace{-0.5cm}
\begin{center}
\includegraphics[width=0.9\textwidth]{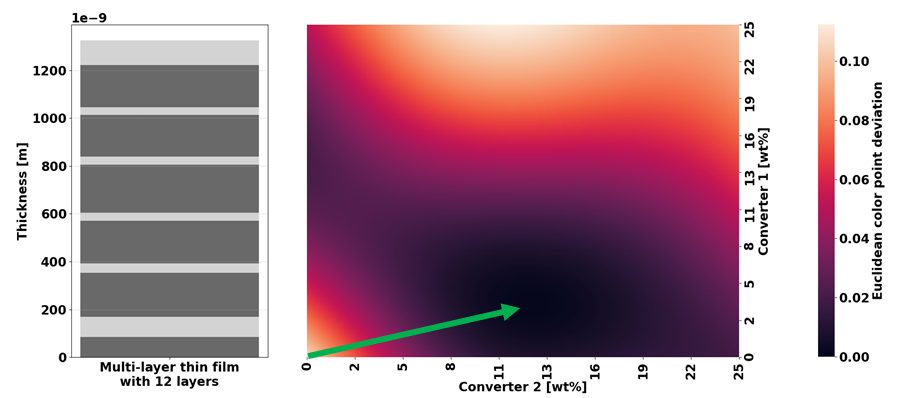}
\caption{(a) Illustration of a multi-layer thin film consisting of two dielectric materials ($\text{TiO}_2$ in light grey and $\text{SiO}_2$ in dark grey). (b) The heatmap of the Euclidean color point deviation over weight percentages of two conversion materials for the multi-layer thin film in (a). The green arrow indicates the possible path of a local optimizer initialized near the origin of the search space to the optimal converter weight percentages.}
\label{fig:convex}
\end{center}
\end{figure}

\begin{figure}[!b]
      \begin{algorithm}[H]
        
        \begin{algorithmic}[1]
          \State Initialize $\mathcal{D}$, $\mathcal{W}$, $0<\alpha<90$, and set a stopping criterion, e.g. a timeout
          \While{stopping criterion not fulfilled}
          \State Suggest (next) $\mathbf{t}$ based on $\mathcal{D}$ and TS-SOO \cite{Bradford2018}
          \State Solve $\mathbf{w} = \text{argmin}_{\boldsymbol{\omega}} \lbrace d ^\alpha \left( 					\boldsymbol{\omega} , \mathbf{t} \right) \rbrace$ based on Downhill-simplex algorithm
          \State Add aquired data point to data set $\mathcal{D} \leftarrow \mathcal{D} \cup 
				\lbrace \left( \mathbf{t} , f ^\alpha (\mathbf{t} \right) \rbrace$
		   and track $\mathcal{W} \leftarrow \mathcal{W} \cup 
				\lbrace \mathbf{w} \rbrace$
          \EndWhile
        \end{algorithmic}
         \caption{Hierarchical optimization approach: Pseudo-code that illustrates the proceeding during optimization. $\mathcal{D}$ denotes the acquired data set of parameters and observations, $\alpha$ denotes the opening angle of the forward cone, and the timeout criterion defines how long the optimization endures. Optionally, the weight percentages are stored in $\mathcal{W}$ for traceability.}
         \label{alg:hierarchical_optimizer}
      \end{algorithm}
      \vspace{-0.3cm} 
\end{figure}

\section{Hierarchical optimization as a product of weights}
\label{problem}

Since hitting a suitable color point of an LED is mandatory for production, we reformulate the problem introduced in section \ref{introduction} as a hierarchical optimization task
\begin{align}
\label{eq:problem}
\max_{\mathbf{t}} \big\lbrace P^\alpha \left( \mathbf{w}, \mathbf{t} \right) \vert \mathbf{w} = \text{argmin}_{\boldsymbol{\omega}} \lbrace d^\alpha\left( \boldsymbol{\omega} , \mathbf{t} \right) \rbrace \big\rbrace.
\end{align}
Therefore, according to domain expert knowledge, we assume that the color point optimization with respect to the conversion materials remains convex for a fixed particular MLTF, if initialized near the origin (see figure \ref{fig:convex}). Equation \eqref{eq:problem} allows to separate the two sets of parameters: The convex color point optimization problem based on the weight percentages $\mathbf{w}$ of conversion materials, and the non-convex power optimization problem including the layer thicknesses $\mathbf{t}$. In this work, a variant of active learning called Bayesian optimization is applied to maximize the power $P^\alpha$ by adapting the thicknesses of the layers of an MLTF. Eventually, a variation of the Downhill-simplex\cite{Nelder1965} optimizer is used to customize the conversion system for each particular MLTF in order to retain the desired color point before evaluating the power. To account for this hierarchical structure, we implement a weighted power as physics-guided real-valued objective function 
\begin{align}
\label{eq:objective}
f ^\alpha ( \mathbf{t} ) = P^\alpha \left( \mathbf{w}, \mathbf{t} \right) 
\cdot W\left(d ^\alpha \left(\mathbf{w} , \mathbf{t} \right) \right)
&\text{, where   }
\mathbf{w} = \text{argmin}_{\boldsymbol{\omega}} \lbrace d ^\alpha \left( \boldsymbol{\omega} , \mathbf{t} \right) \rbrace  \\
&\text{ and   }
W \left( d^\alpha \left(\mathbf{w} , \mathbf{t} \right) \right) = \text{exp} \left( a \cdot [ d  ^\alpha \left(\mathbf{w} , \mathbf{t} \right) ] ^b \right),
\label{eq:weighting}
\end{align}
which is maximized via Thompson-sampling single-objective Bayesian optimization\cite{Bradford2018} (TS-SOO) in this work. Here, $\mathbf{w}$ is a solution to the nested color point optimization in problem \eqref{eq:problem}. As explained, the weight percentage parameters $\mathbf{w}$ are not tuneable by the Bayesian optimizer directly. 
However, for given thicknesses $ \mathbf{t}$ of an MLTF, the color point deviation $d^\alpha(\mathbf{w} , \mathbf{t})$ may be bounded from below with regard to the weight percentages $\mathbf{w}$. This means that the target color point is not reachable for a given MLTF. In such cases, the comparison of power values for different MLTFs at different color points becomes invalid due to the Stokes shift. Here, $W(\cdot)$ allows to guide the course of optimization towards suitable MLTFs: The weighting punishes excessive deviations from the target color point by decreasing the objective, although the power of an MLTF may be high. In practice, more weighting functions may be introduced to account for various conflicting or competing effects during LED development.\par
We compared the TS-SOO to the Thompson-sampling efficient multi-objective optimization\cite{Bradford2018} (TS-EMO). Here, a high forward power $P^\alpha$ and a low color point deviation $d^\alpha$ are considered as (competing) real-valued objectives. Namely, TS-EMO directly searches for joint parameter vectors $\left(\mathbf{w}, \mathbf{t} \right)$ to entry-wise maximize
\begin{align}
\label{eq:joint}
\mathbb{R}^2 \times \mathbb{R}^{T} &\rightarrow \mathbb{R}^2 \\
\nonumber
\left(\mathbf{w}, \mathbf{t} \right) &\mapsto \left( P^\alpha\left(\mathbf{w}, \mathbf{t} \right), - d^\alpha\left( \mathbf{w} , \mathbf{t} \right) \right),
\end{align}
where the aforementioned Stokes shift renders the involved objectives to be competitive in nature. The minus sign of the second entry of equation \eqref{eq:joint} reflects the intention to minimize the color point deviation. Note that the hierarchical structure of the engineering optimization problem is not longer represented in the equation.

\begin{figure}[!t]
\vspace{-0.5cm}
\includegraphics[width=0.99\textwidth]{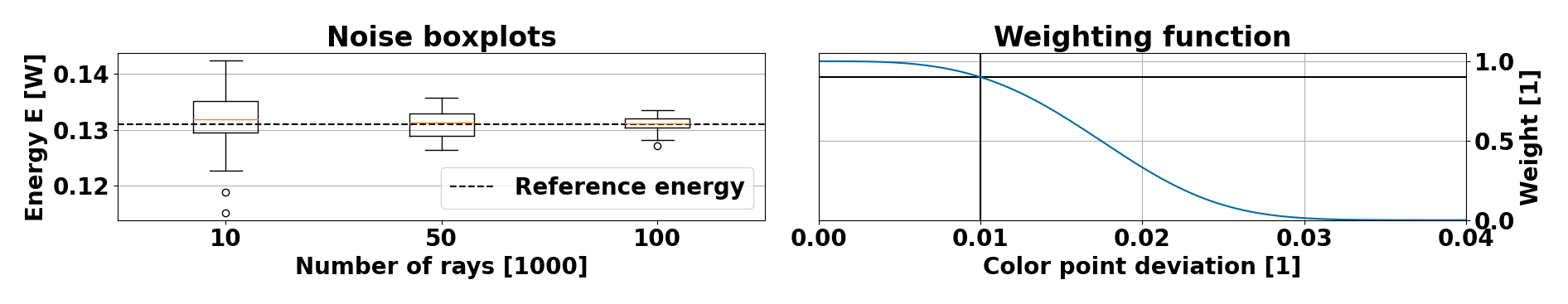}
\caption{(left) The ray tracing for the reference design based on different numbers of traced rays ($10 \cdot 10^3$, $50 \cdot 10^3$ and $100 \cdot 10^3$) is started $15$ times. The conversion material weight percentages remained unchanged. As expected, the corresponding boxplots indicate that the noise level decreases with an increasing number of rays traced. The horizontal dashed line indicates the reference power obtained with $10^6$ rays. This noise effect holds for the estimated color point $\left( C_x, C_y \right)$, too. (right) The weighting function \eqref{eq:weighting} for $a=-634914.5425$ and $b=3.3900$. The dark lines emphasize the condition $W(0.01) = 0.9$.}
\label{fig:implementation}
\end{figure}

\section{Implementation}
For applications like headlights, we set $\alpha=25^\circ$ and set $d^\alpha(\mathbf{w} , \mathbf{t}) = 0.005$ as a preliminary upper bound of the color point deviation from the required target color point $\mathbf{C} = \left( \nicefrac{1}{3}, \nicefrac{1}{3} \right)$ of a white LED. This enables us to universally solve for the parameters $ \left( a, b \right)$ such that the empirically derived conditions
\begin{align}
\label{eq:weight}
W(0.005) &= 0.99 \\ \nonumber W(0.010) &= 0.90
\end{align}
hold, yielding $a=-634914.5425$ and $b=3.3900$. The resulting weighting function is illustrated in figure \ref{fig:implementation} (a). The reference design refers to the special case $t_1 = ... = t_{T} = 0.0$. In general, we propose to formulate hierarchical, competing objectives as a product of weights and a central figure of merit, e.g. the power. Conceptually, this enforces strong (AND-)conditions for all constraints explained by the weights. In other words, a high figure of merit is only valuable if all constraints are fulfilled. On the other hand, this formulation preserves continuity of the objective function which makes it approximable with Gaussian processes \cite{Rasmussen2005}. As the black-box function \eqref{eq:objective} is based on a ray tracing simulation of $25 \cdot 10^4$ rays that takes several minutes ($\approx 4 \text{ [min]}$) to be evaluated and features non-linearity and non-convexity, the maximization is conducted via an active learning appoach, the aforementioned TS-SOO. This variant of Bayesian optimization is employed to optimize expensive-to-evaluate engineering and chemical problems \cite{Schweidtmann2018,Amar2019,Clayton2020}. Moreover, as illustrated in figure \ref{fig:implementation} (b), the evaluation of the objective function provides noisy samples due to the conducted ray tracing. As explained in algorithm \ref{alg:hierarchical_optimizer}, in each optimization iteration $n$ a thickness vector $\mathbf{t}^n$ is suggested using TS-SOO. For this MLTF, a variant of the Downhill-simplex algorithm solves the convex color point optimization nested in statement \eqref{eq:problem}, yielding $\mathbf{w}^n$. The acquired data point $\left( \mathbf{t}^n , f ^\alpha \left(\mathbf{t}^n \right) \right)$ is added to the data set $\mathcal{D}$. This data set is used to update the global surrogate model --- implemented as a Gaussian process --- based on which the next thickness vector $\mathbf{t}^{n+1}$ is derived until a predefined stopping criterion is fulfilled, e.g. a timeout or a maximum number $N \geq n$ of iterations. Basically, TS-EMO follows the same optimization routine, but directly suggests joint parameter vectors $\left( \mathbf{w}^n, \mathbf{t}^n \right)$ to solve the multi-objective optimization problem \eqref{eq:joint}. For TS-SOO as well as TS-EMO, we set $T=12$ and allow the thicknesses to vary between $10 \text{ [nm]}$ and $200 \text{ [nm]}$ for each layer. Moreover, both conversion material weight percentages are adoptable between $0 \text{ [wt\%}]$ and $25 \text{ [wt\%}]$. After the objective function \eqref{eq:objective} or \eqref{eq:joint} was optimized via TS-SOO or TS-EMO, a naive Downhill-simplex algorithm \cite{Nelder1965} is conducted, respectively. Via not more than $25$ optimizer steps the thickness parameters and conversion material weight percentages are jointly fine-tuned. Each of these steps takes $8-9  \text{ [min]}$ and is based on $5 \cdot 10^{5}$ traced rays per simulation to evaluate the respective objective function. During the local refinement, the acceptable upper bound for the color point deviation in equation \eqref{eq:weight} is narrowed down from $0.005$ to $0.002$, which is practically required for most applications.

\begin{figure}[!t]
\vspace{-0.5cm}
\includegraphics[width=0.99\textwidth]{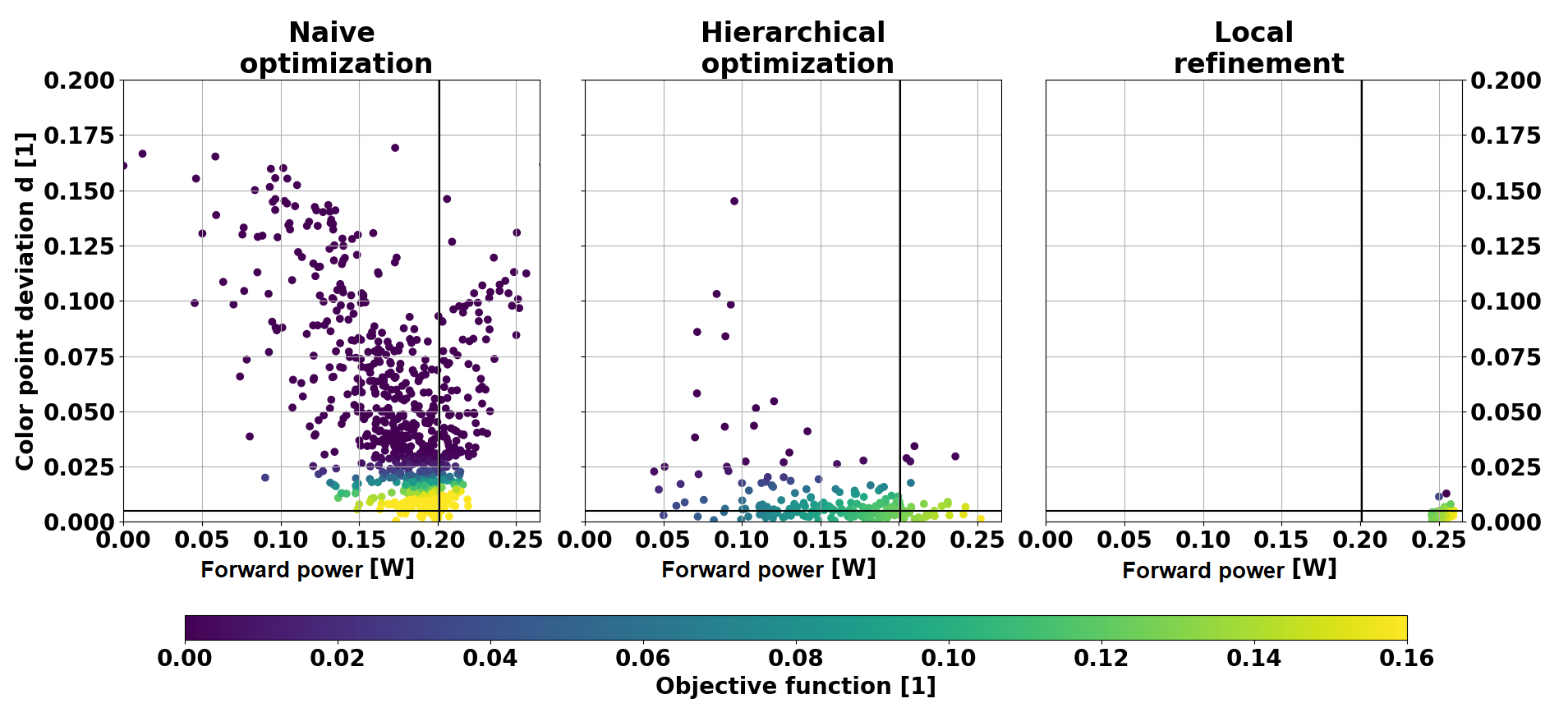}
\caption{Color point deviation $d^\alpha$ over power $P^\alpha$ after joint (left, TS-EMO), hierarchical (middle, TS-SOO) and local (right, Downhill simplex) optimization for $\alpha=25^\circ$. The horizontal black lines denote the color point deviation of $0.005$ points, where $W\left( 0.005 \right) = 0.99$ holds. The vertical black lines denote the power of the reference design without multi-layer thin film. For each sample in these plots, the color map reflects the objective function values obtained by equation \eqref{eq:objective}.} 
\label{fig:results}
\end{figure}

\section{Results}
In this section, we present the results of our investigations. First, we demonstrate that MLTFs can increase the directionality of white LEDs, which may seem counterintuitive beforehand. Second, we give an explanation of the underlying physical effects and discuss how we can make the latter observable in the spectra. The results are summarized in table \ref{tab:results}. Here, we report the MLTFs suggested by TS-SOO and TS-EMO that provided the highest power in $\pm25^\circ$, while exhibiting a color point deviation lower than $0.002$ points --- which is an acceptable deviation for the most consumer products in practice. As mentioned, a joint local optimization of both, thicknesses and conversion material weight percentages is conducted after the global Bayesian optimization. Unsurprisingly, the total power of the white LED decreases for all considered MLTFs due to absorption losses. However, more power is available at particular forward angles of interest. This directionality increase is of value for applications like automotive headlamps or projection, where non-forward light does not contribute.

\begin{figure}[!t]
\vspace{-0.5cm}
\includegraphics[width=0.99\textwidth]{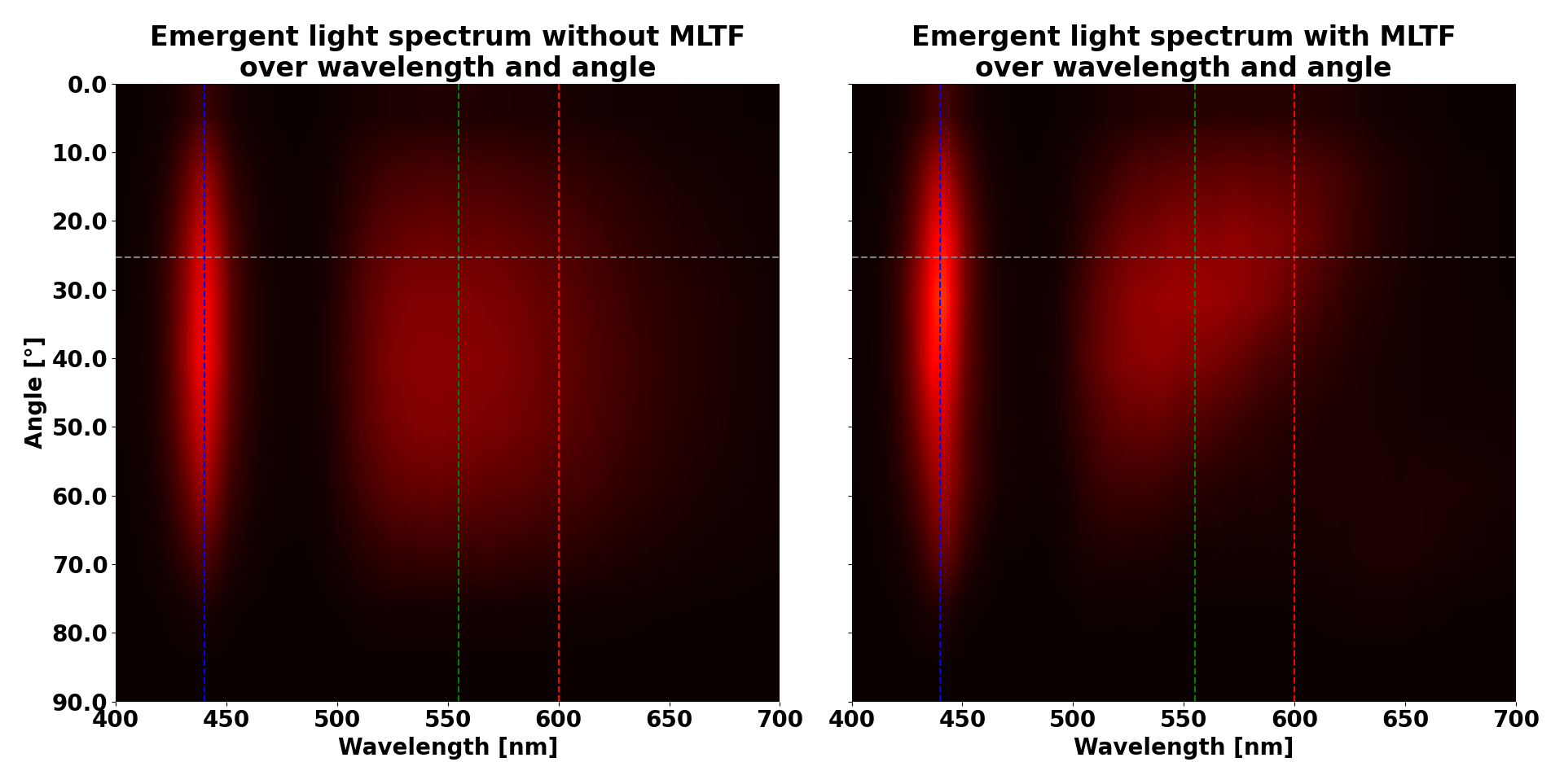}
\caption{(left) Common angular and spectral distribution of relative power of a white LED without a multi-layer thin film. (right) Spectrum of a white LED with a multi-layer thin film that contains $28.9 \%$ more relative power compared to the reference design in forward direction. In each case, the blue chip wavelength ($440\text{ [nm]}$) and the green and red wavelengths ($555\text{ [nm]}$) and $600\text{ [nm]}$) of the conversion materials are indicated by vertical dashed lines. The horizontal grey line indicates the $\pm 25^\circ$-forward direction. The color map for the heat maps coincides with figure \ref{fig:photonpingpong} (right).}
\label{fig:spectra}
\end{figure}

\begin{table*}[!b]
\centering
\footnotesize
\makebox[\linewidth]{
\begin{tabular}{ | P{1.5cm} | P{2cm} || P{2cm} || P{2cm} | P{2cm} | } 
 \hline
  & $\alpha$ \textbf{ } $[^\circ]$ & $25$ & $45$ & $90$\\
 \hline
  \hline
 $P^\alpha \text{ [W]}$ & \pbox{25cm}{\vspace{5pt} Reference \\ TS-SOO \\ TS-EMO \vspace{5pt}} & 
 \pbox{20cm}{$0.201$ \\ $\mathbf{0.259}$ \\ $0.210$} & \pbox{20cm}{$0.559$ \\ $0.640$\\ $0.569$} & \pbox{20cm}{$1.052$ \\ $0.985$ \\ $0.978$} \\
  \hline
 $d^\alpha$ \textbf{ } $[1]$ & \pbox{20cm}{ \vspace{5pt} Reference \\ TS-SOO \\ TS-EMO \vspace{5pt}} & 
 \pbox{20cm}{$0.0015$ \\ $\mathbf{0.0007}$ \\ $0.0025$} & \pbox{20cm}{$0.0071$ \\ $0.0230$ \\ $0.0262$} & \pbox{20cm}{$0.0146$ \\ $0.0275$ \\ $0.0281$} \\
 \hline
\end{tabular}
}
\caption{Optimization results regarding color point deviation $d^\alpha$, and power $P^\alpha$ of algorithm \ref{alg:hierarchical_optimizer}, and TS-EMO for $\alpha = 25^\circ$, and after local refinement. The designs featuring a multi-layer thin film are compared with the reference design, where only a color point optimization is conducted. In addition to the direct optimization objectives regarding $\pm 25^\circ$, we also report the color point deviations and energies for $\pm 45^\circ$ and $\pm 90^\circ$ observed for the respective optimized MTLFs.}   
\label{tab:results}
\end{table*}

\subsection{Increase of directionality}
The reference design provides $P^\alpha\left(\mathbf{w},\mathbf{t}=\mathbf{0}\right)=0.201 \text{ [W]}$ after adopting the weight percentages $\mathbf{w}$, while the deviation between the LED's and the target color point is given by $d^\alpha \left( \mathbf{w}, \mathbf{t} = \mathbf{0} \right) = 0.00151 < 0.002$. After optimizing an MLTF's thickness vector $\mathbf{t}$ using algorithm \ref{alg:hierarchical_optimizer}, the forward power of the LED is increased by $25.4 \%$ to $P^\alpha \left( \mathbf{w}, \mathbf{t} \neq \mathbf{0} \right) = 0.252 \text{ [W]}$. The Euclidean color point deviation for this design is given by $0.0014< 0.002$. Notably, using TS-EMO to solve the corresponding original multi-objective optimization problem \eqref{eq:joint} achieved only almost $5\%$ more forward power while keeping the color point deviation below the required value of $0.002$. This is not surprising, as TS-EMO is not supposed to be used for more than eight parameters\cite{Bradford2018}. As illustrated in the very right plot in figure \ref{fig:results}, the local refinement increased the power in forward direction by an additional $3.5 \%$ relative to the reference design. Thus, yielding $0.259  \text{ [W]}$ or $28.9 \%$ more light compared to the reference design, while maintaining an acceptable color point deviation of $0.0007 < 0.002$. Notably, starting a local refinement with the joint parameters provided by the TS-EMO implementation, either the color point deviation could not be reduced under $0.002$ or the forward power increase remained below $10.0\%$. Due to the high computational effort associated with ray tracing simulations, we restarted each optimization procedure---based on TS-SOO and TS-EMO--- only three times for $48\text{ [hours]}$ each. The reported data corresponds to the best run regarding the objective function for TS-SOO and TS-EMO, respectively. Because the relative and absolute trends of the results of these experiments appeared to be consistent and showed no unexpected anomalies, they were not evaluated in detail. The Pareto fronts illustrated in figure \ref{fig:results} plot the (forward) power against the color point deviation for both, joint optimization using TS-EMO and hierarchical optimization using TS-SOO. As the time horizont was fixed, the different numbers of samples are explained by the implementation of the joint ($719$ samples) and hierarchical ($143$ samples) optimization approach: For each thickness vector $\mathbf{t}$ suggested by TS-SOO, the solving of the nested color point optimization \eqref{eq:objective} based on ray tracing simulations takes about $20 \text{ [min]}$. Contrariwise, the evaluation of a joint parameter vector $\left(\mathbf{t}, \mathbf{w}\right)$ requires only one ray tracing simulation and thus takes about $4\text{ [min]}$. The comparison between the left and middle Pareto fronts in figure \ref{fig:results} indicate that TS-SOO circumvents MLTFs that lead to high color point deviations, as those are punished via multiplicative weights \eqref{eq:weighting}. In contrast, TS-EMO is not implicitly informed about the engineering structure of the problem via the objective function \eqref{eq:joint}. Namely, an increase in forward power at the cost of color point deviation is of no value for specific LED applications. Therefore, most MLTFs suggested by TS-EMO indeed achieve the same or even higher forward power values compared to TS-SOO, but bring along an unacceptable color point deviation significantly above $0.005$.

\begin{figure}[!t]
\vspace{-0.5cm}
\includegraphics[width=0.99\textwidth]{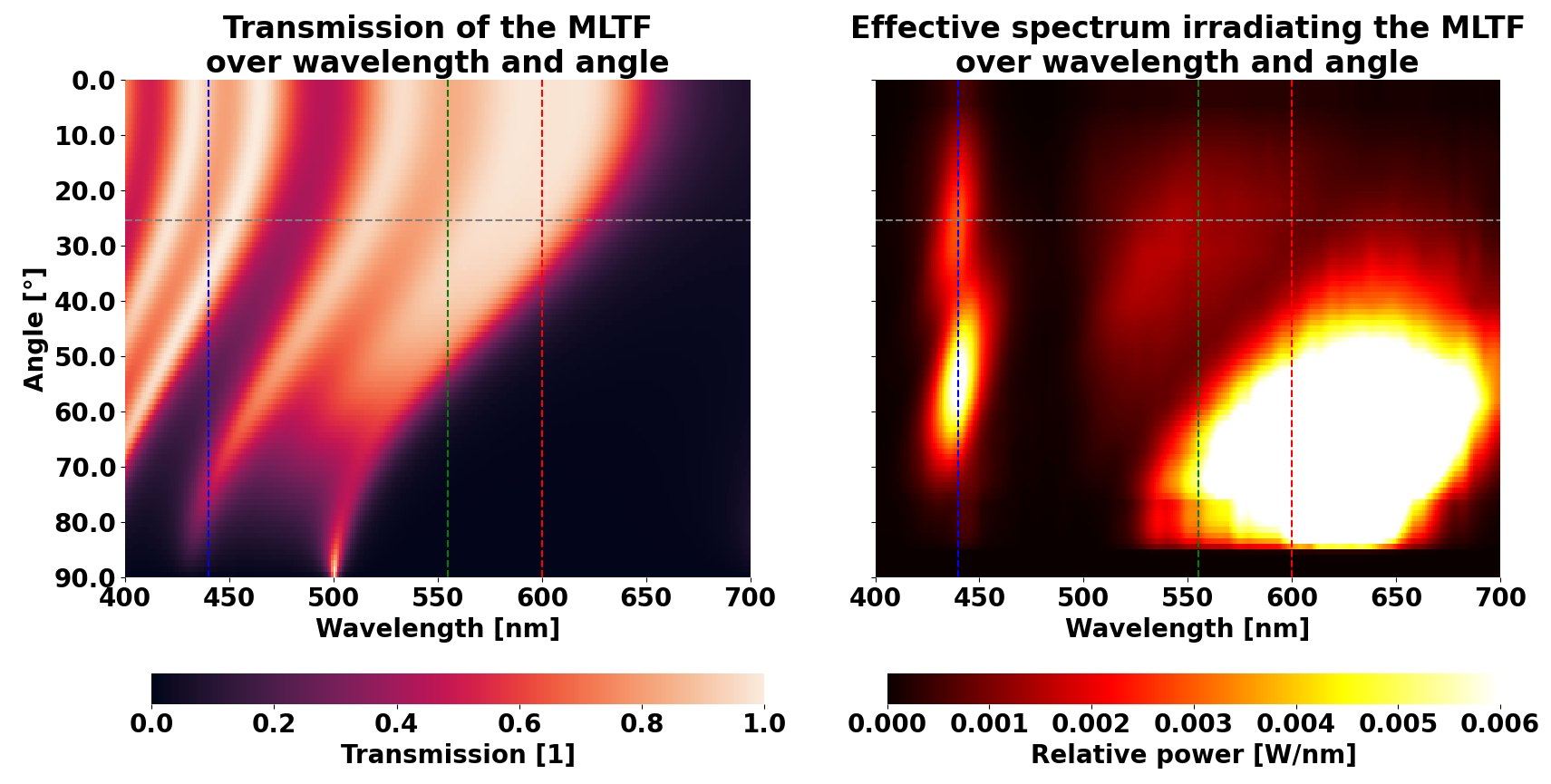}
\caption{(left) Transmission of a multi-layer thin film that exhibits $28.9 \%$ more light compared to the reference design in forward direction. (right) The hypothetical, effective spectrum of light, which exits the conversion system and irradiates the multi-layer thin film. In both cases, the vertical dashed lines indicate the wavelengths contributing to the LED spectrum and the horizontal lines denote the separation between forward and non-forward direction.}
\label{fig:photonpingpong}
\end{figure}

\subsection{Ping pong with light rays}
The results of the previous section allow to deduce an explanation of the mechanism of white light directionality using an MLTF. We can deduce a hypothetical effective spectrum which irradiates the  MLTF. Therefore, we conduct pixel-wise division of the observed spectrum in figure~\ref{fig:spectra} (b) through the spectral and directional transmission of the optimized MLTF in figure \ref{fig:photonpingpong} (a). This hypothetical spectrum is illustrated in figure \ref{fig:photonpingpong} (b) and indicates how much relative power is required to impinge on the MLTF from the conversion system for each angle and wavelength, in order to explain the physically detected spectrum that irradiates the ambient air after traversing the MLTF. The transmission describes the probability of a light ray to pass the MLTF, depending on its wavelength and angle of incidence. It was computed by a parallelized version of the TMM package\cite{Byrnes2020} provided by Luce et al.\cite{Luce2021}. Here, the light-injecting substrate (conversion system) is represented as silicone ($\text{SiO}_2$) of infinite thickness. Air of infinite thickness defines the ambient environment. Aside from the relative power peak around the chip wavelength of $440 \text{ [nm]}$, another two contributions of relative power at $555 \text{ [nm]}$ and $600 \text{ [nm]}$ appear as a single broad peak in the spectra of figure \ref{fig:spectra}. These green and red wavelength contributions correspond to the conversion material emissions, respectively. As expected, the transmission of the MLTF for all of these wavelengths is low for inconvenient, large beam angles. Thus, non-forward light rays are trapped in the LED package until physical interactions change their directional properties such that they are likely to escape or the light rays vanish optically due to non-radiative thermal effects. We suppose that this phenomenon causes the directionality enhancement and refer to it as \textit{ray ping pong}. Here, the MLTF on top of the conversion system directly influences the emitted spectrum over angle and wavelength of the LED. The MLTF does not only function as an angle selective filter that reflects non-forward light rays back into the LED package, but also balances the statistics of emitted blue, green and red rays in order to appear as white light of a specific color. In other words, the MLTF exploits the process of ray ping pong to reach a statistical equilibrium of emitted rays that is advantageous regarding directionality and color point of the outcoupled light. Notably, the characteristic transmission pattern of an MLTF over wavelength and angle of incidence (see figure \ref{fig:photonpingpong}) not only depends on the layer thicknesses, but also the constituent materials. Studying table \ref{tab:results} implies that the MLTF decreases the total efficacy of a white LED corresponding to $\alpha = 90^\circ$: Rays are trapped in the LED package and are thus more likely to further interact with the LED materials instead of escaping into the ambient air. Such interactions may include photon recycling or scattering, but also non-radiative effects like thermal losses of rays absorbed by the conversion materials. As mentioned, for applications like head lamps only power emitted in forward direction $\left( \alpha \ll 90^\circ \right)$ is usable. In this case, any increase in directionality obviously outweighs a (moderate) drop of global efficacy.
\newpage
\subsection{Limits of directionality increase}
\begin{wrapfigure}{R}{0.6\textwidth}
\vspace{-1.5cm}
  \begin{center}
    \includegraphics[width=0.59\textwidth]{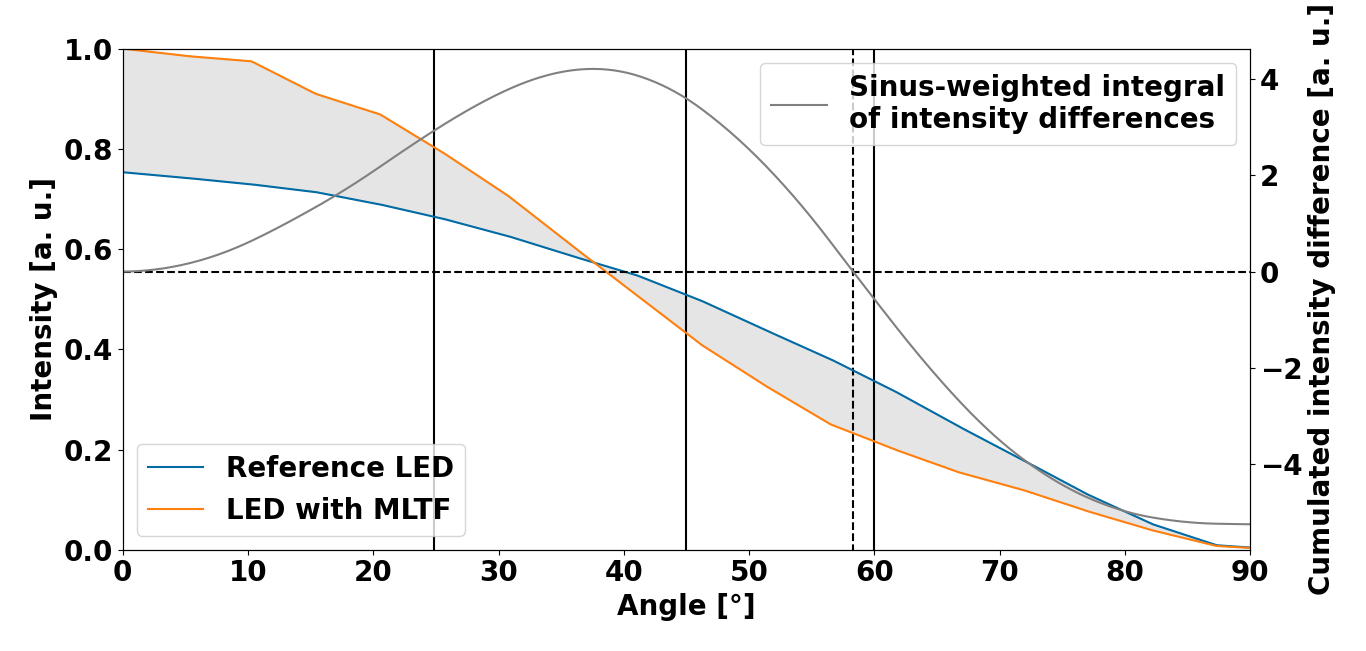}
  \end{center}
  \caption{The normalized intensity of an LED with and without a multi-layer thin film (MLTF) is denoted on the left axis. The sinus-weighted integral over the intensity difference (grey area) --- the so-called \textit{cumulated intensity difference} --- on the right axis indicates that a directionality increase can be achieved up to $58.4^\circ$. However, the peak of directionality increase is observed between $35^\circ$ and $40^\circ$.}
  \label{fig:intensity}
\end{wrapfigure}
The previous section allows us to understand the constraints of our approach. Although no explicit optimizations for angles different from the application-specific $25^\circ$ were conducted for this work, our investigations regarding the cumulated intensity difference of figure \ref{fig:intensity} indicate that an increase in forward power is possible up to almost $60^\circ$. Namely, for angles $\alpha>58.4^\circ$, the MLTF would need to re-direct light rays of unsuitable angles between $\alpha$ and $90^\circ$. Due to the Lambertian radiation characteristics of LEDs, the quantity of such rays may be too low to outweigh non-radiative effects like thermal losses, which are statistically enforced by the MLTF for each ray. Thus, the deposition of an MLTF may not be suitable for applications that leverage light with a broad range of incidence angles, here larger than $58.4^\circ$.
\section{Conclusions}
Increasing the (forward) power of a white LED while still maintaining its color point is a hierarchical multi-objective optimization problem: The Stokes shift renders these objectives to be competitive by nature, because a naive increase in power may turn the emitted light bluish instead of pure white. Since the color point is a strict requirement for many applications, a higher forward power of an LED at the cost of color changes is utterly undesired. In this work, we enter into the competition between color point and power with a Bayesian optimization approach that optimizes a physics-guided, weighted objective function. Hereby, weights continuously implement hard constraints and thus preserve approximability via Gaussian processes. The reported results indicate that multi-layer thin films on top of white LEDs can increase the light directionality. The epitaxial deposition of such multi-layer thin films is only implying low additional expenditure and directly applicable for mass-production of many optical semiconductors. Our analyses reveal that a carefully designed multi-layer thin film functions as an angle and wavelength sensitive filter: The filter statistically balances emitted rays of different wavelengths to meet the color point. In addition, it traps rays that would exit the LED at large angles in order to implicitly enforce their forward (re-)emission. To summarize, the proposed objective function guides the optimization towards a multi-layer thin film that leverages statistical ray ping pong to enforce favorable properties of the spectrum radiated by the LED. Thus, we shine a light on the previously enigmatic effect causing the counterintuitive increase of white light directionality using a multi-layer thin film. 
\section*{Material, Data, and Code Availability}
The optical models and related (commercial) software that support the findings of this study are available from OSRAM Opto Semiconductors GmbH but restrictions apply to the availability of these items, which were used under license for the current study, and so are not publicly available. The generated and analysed data as well as any code-related information is included in this manuscript or already published by third parties.

\bibliography{BayesOptimizationDirectionality}

\section*{Acknowledgements}
The authors would like to thank the engineers and scientists, especially Daniel Gr\"unbaum, at OSRAM Opto Semiconductors GmbH in Regensburg for their exhaustive assistance and for providing the required optical models and computational resources.

\section*{Author contributions statement}
H.W. developed the physics-guided objective function, wrote and ran the code, devised the experiments and edited the manuscript. C.W. and L.K. interpreted the acquired data in terms of the ray ping pong and contributed application-specific knowledge about white light. R.B. analysed the results and advised how to avoid the Stokes shift. A.L. provided code to evaluate and discuss multi-layer thin films. S.S. provided the calibrated optical model of a white LED to conduct ray tracing simulations. M.L.S and E.W.L made substantial conceptional and editorial contributions to this manuscript. All authors reviewed the manuscript \textit{and played a round of ray ping pong}.

\end{document}